\documentclass[prd,twocolumn,floatfix,nofootinbib,superscriptaddress]{revtex4}
 \usepackage[utf8]{inputenc}
\usepackage{graphicx}
\usepackage{epsfig}
\usepackage{bm}
\usepackage{amssymb}
\usepackage{float}
\usepackage{amssymb}
 
\usepackage{amsmath}
\usepackage{dcolumn}
\usepackage{tensor}
\usepackage{dcolumn}
\usepackage{tensor}
\usepackage{cancel}
\usepackage[colorlinks]{hyperref}
\usepackage[usenames,dvipsnames]{color}
\hypersetup{
     breaklinks=true,
    pdfstartview={FitH},    % fits the width of the page to the window
    colorlinks=true,       % false: boxed links; true: colored links
    linkcolor=blue,          % color of internal links
    citecolor=red,        % color of links to bibliography
    filecolor=magenta,      % color of file links
    urlcolor=blue,           % color of external links
    anchorcolor=green,      % Color for anchor text
    linktocpage=true
}

\def\b{ \beta}
\def\g{ \gamma}
\def\d{ \delta}
\def\m{ \mu}
\def\n{ \nu}

\def\r{ \rho}
\def\s{ \sigma}

\def\b{ \beta}
\def\g{ \gamma}
\def\d{ \delta}
\def\m{ \mu}
\def\n{ \nu}

\def\r{ \rho}
\def\k{ \kappa}

\def\be{\begin{equation*}}
\def\ee{\end{equation*}}

\usepackage{amsmath,latexsym}
\usepackage{placeins}
\usepackage[toc,page]{appendix}
\newcommand{\mathsym}[1]{{}}
\newcommand{\unicode}[1]{{}}

\newcommand{\bea}{\begin{eqnarray}}
\newcommand{\eea}{\end{eqnarray}}
\newcommand{\beaa}{\begin{eqnarray*}}
\newcommand{\eeaa}{\end{eqnarray*}}

\def\arxiv{http://arxiv.org/abs}
\begin{document}
\title{Building cubic gravity with healthy and viable scalar and tensor 
perturbations}

\author{Petros Asimakis}
\affiliation{Department of Physics, National Technical University of Athens, 
Zografou Campus GR 157 73, Athens, Greece}

\author{Spyros Basilakos}
\affiliation{National Observatory of Athens, Lofos Nymfon, 11852 Athens, 
Greece}
\affiliation{Academy of Athens, Research Center for Astronomy and Applied 
Mathematics, Soranou Efesiou 4, 11527, Athens, Greece}
\affiliation{School of Sciences, European University Cyprus, Diogenes 
Street, Engomi, 1516 Nicosia, Cyprus}

\author{Emmanuel N. Saridakis}
\affiliation{National Observatory of Athens, Lofos Nymfon, 11852 Athens, 
Greece}
\affiliation{CAS Key Laboratory for Researches in Galaxies and Cosmology, 
Department of Astronomy, University of Science and Technology of China, Hefei, 
Anhui 230026, P.R. China}
 \affiliation{Departamento de Matem\'{a}ticas, Universidad Cat\'{o}lica del 
Norte,  Avda. Angamos 0610, Casilla 1280 Antofagasta, Chile}

\begin{abstract}  
We investigate sufficient conditions 
under which cubic gravity is healthy and viable at the perturbation level. We 
perform a detailed analysis of the scalar and tensor perturbations. We impose 
the requirement that the 
two scalar potentials, whose ratio is the   post-Newtonian parameter $\gamma$, 
should   deviate only minimally form general relativity. 
Additionally, concerning   tensor perturbations we  impose satisfaction of 
the  LIGO-VIRGO 
and Fermi Gamma-ray Burst observations, and thus we result to a 
gravitational-wave equation with gravitational-wave speed equal to the speed of 
light, and where the only deviation from general relativity appears in the 
dispersion relation. Furthermore, we show that cubic gravity exhibits an 
effective Newton's constant that depends on the model parameter, on the 
background evolution, and on the wavenumber scale. Hence, by requiring its 
deviation from the standard Newton's constant to be within observational bounds 
we extract the constraints on the single coupling parameter $\beta$. 
\end{abstract}
%\pacs{98.80.-k, 04.50.Kd}

\maketitle

\section{Introduction}
 
Theories of gravity with higher-order invariants  arise naturally 
as an effective description of a complete String Theory \cite{Gross:1986mw}, 
and since they can improve the renormalizability properties of general 
relativity \cite{Stelle:1976gc} they have attracted the interest of the 
literature \cite{Addazi:2021xuf}. On the other hand, one may have an additional 
motivation of cosmological origin 
\cite{CANTATA:2021ktz,Nojiri:2006ri,Clifton:2011jh}, since when applied at a 
cosmological framework such theories may lead to new effective sectors of 
gravitational origin, that can drive inflation or late-time acceleration, or 
alleviate the $H_0$ and $S_8$ cosmological tensions 
\cite{Abdalla:2022yfr,DiValentino:2022uvj}.

In order to construct higher-order theories of gravity one  starts
from the Einstein-Hilbert Lagrangian and includes  extra higher-order  terms, 
such as in 
 $f(R)$ gravity 
\cite{Starobinsky:1980te,Capozziello:2002rd,DeFelice:2010aj},
 in  $f(G)$
gravity \cite{Nojiri:2005jg, DeFelice:2008wz},  in 
Lovelock
gravity \cite{Lovelock:1971yv, Deruelle:1989fj},  etc. Alternatively, but not 
equivalently, one can construct higher-order theories of gravity in the 
torsional formulation, resulting in  $f(T)$ gravity
\cite{Cai:2015emx,Linder:2010py, Chen:2010va}, in $f(T,T_G)$ 
gravity
\cite{Kofinas:2014owa,Kofinas:2014daa}, in $f(T,B)$ gravity 
\cite{Bahamonde:2015zma,Bahamonde:2016grb}, etc.

 In general such gravitational constructions involve extra degrees of freedom, 
which may be problematic, giving rise to various pathologies, such as ghost and 
Laplacian instabilities. Hence, one needs to focus on subclasses of these 
theories that are free from pathologies. We stress here that 
this has to hold around all backgrounds, and at all orders in perturbation 
theory, since a well-behaved background evolution does not necessarily 
guarantee  well-behaved perturbations (as for instance was the case in the 
initial versions of Ho\v{r}ava-Lifshitz   
\cite{Charmousis:2009tc,Bogdanos:2009uj}, of  new nonlinear massive 
gravity\cite{Hinterbichler:2011tt}, of entropic-force dark energy  
\cite{Basilakos:2014tha}, etc).

Nevertheless, theoretical consistency is a necessary but not sufficient 
condition for the acceptance of a particular theory, since observational and 
experimental viability should also be obtained. Therefore, every theory should 
satisfy the bounds acquired by Solar System experiments \cite{Will:2014kxa}, as 
well as be in agreement with   various datasets of cosmological observations, 
such as  Supernovae Type I (SNIa), Baryonic Acoustic Oscillations (BAO), direct 
Hubble constant measurements with cosmic chronometers (CC),  
Cosmic Microwave Background (CMB) shift temperature and polarization, 
 redshift space distortion ($f\sigma_8$) and Large Scale Structure measurements 
\cite{Cai:2009zp}, etc. Finally, since modified theories of gravity may predict 
gravitational wave speed $c_{_{GW}}$ different than the speed of light $c$, one 
must guarantee that she can satisfy   the LIGO-VIRGO 
\cite{LIGOScientific:2017vwq} and Fermi Gamma-ray Burst Monitor 
\cite{Goldstein:2017mmi} observations, which 
require  $|c_{_{GW}}/c-1|\leq 4.5\times 
10^{-16}$ \cite{Monitor:2017mdv}.

One interesting class of higher-order gravity is the one that contains as a 
Lagrangian the cubic combination of curvature terms $P$  \cite{Bueno:2016xff}, 
which was then extended to  $f(P)$ gravity 
\cite{Erices:2019mkd}. In  \cite{BuenoCano,Mann01,Mann02,Ghodsi:2017iee} the 
authors reduced the possible coefficients in order to obtain  second-order 
field 
equations that allow for  spherically symmetric black hole solutions, while in 
 \cite{Edelstein01} an  extra cubic correction, which is trivial for a 
spherically symmetric metric, was added in order to lead to second-order field 
equations in a cosmological background too. The resulting cubic and $f(P)$ 
gravities prove  to have interesting cosmological applications, and thus they 
have 
attracted  extensive investigation
\cite{Erices:2019mkd,Bueno:2016xff,Mehdizadeh:2019qvc,Mir:2019rik,
Burger:2019wkq,KordZangeneh:2020qeg,Marciu:2020ysf, Arciniega:2018tnn,
Cisterna:2018tgx,Quiros:2020uhr,Marciu:2020ski,Konoplya:2020jgt,
BeltranJimenez:2020lee,
Bhattacharjee:2021nfx,Giri:2021amc,Sardar:2021blt, Asimakis:2021yct,
Marciu:2022wzh,Bueno:2022res,Anastasiou:2022pzm}.
 
However, despite the significant research on cubic gravity, the 
scalar and tensor perturbation analysis has not been performed. Thus, in the 
present work we are interested in performing such an analysis, and additionally 
we desire to extract conditions on the model parameters that allow for 
healthy and viable theories at the perturbation level. The plan of the work is 
as follows: In Section \ref{model} we present cubic gravity and we provide the 
basic requirement in order to have well-defined cosmological behavior at the 
background level. Then in Section \ref{Perturbationanalysis} we perform a 
detailed scalar and tensor perturbation analysis, extracting the conditions 
corresponding to absence of instabilities as well as to gravitational-wave 
speed equal to the speed of light. Finally, in Section \ref{Conclusions} we 
summarize and conclude.

\section{Cubic gravity}
\label{model}

In this section we briefly review cubic modified gravity. This theory of 
gravity is based on adding corrections to the action of General Relativity, 
constructed from cubic combinations of the Riemann tensor.  A general such 
combination is \cite{Lovelock:1971yv}
\begin{eqnarray}\label{P}
&&
\!\!\!\!\!\!\!\!\!\!\!\!
P=\b_1 
{{{R_{\m}}^{\r}}_{\n}}^{\s}{{{R_{\r}}^{\g}}_{\s}}^{\d}{{{R_{\g}}^{\m}}_{\d}}^{\n
}+\b_2 
R_{\m\n}^{\r\s}R_{\r\s}^{\g\d}R^{\m\n}_{\g\d}
\nonumber\\
&&
+\b_3 
R^{\s\g}R_{\m\n\r\s}{R^{\m\n\r}}_{\g}+\b_4 
R R_{\m\n\r\s}R^{\m\n\r\s}\nonumber\\
&&
+\b_5 R_{\m\n\r\s}R^{\m\r}R^{\n\s}+\b_6 
R^{\n}_{\m}R^{\r}_{\n}R^{\m}_{\r}\nonumber\\
&&+\b_7 
R_{\m\n}R^{\m\n}R+\b_8 R^3,
\end{eqnarray}
where $\beta_i$'s are eight coefficients. Adding the above invariant in the 
Einstein-Hilbert Lagrangian, we can write the action
\begin{equation}
 \label{GB1}
S=\int d^4 x\sqrt{-g}\left[\frac{1}{2\kappa}\left(R-2\Lambda\right) + \alpha 
P+\mathcal{L}_{m}\right] ,
\end{equation}
 where $\alpha$ is a possible coupling parameter,  $\kappa=8\pi G$ is the 
gravitational constant, and where for completeness  we have also added the 
cosmological constant $\Lambda$, as well as the matter Lagrangian 
$\mathcal{L}_{m}$.

Varying the action with respect to the metric $g_{\mu\nu}$ we obtain the 
general field equations, namely \cite{Erices:2019mkd}. 
\begin{equation}\label{eom}
G_{\mu\nu}+\Lambda g_{\mu\nu}=\kappa\left(T_{\mu\nu}+\alpha H_{\mu\nu}\right),
\end{equation}
with $T_{\mu\nu}=-\frac{2}{\sqrt{-g}}\frac{\delta(\sqrt{-g}L_{m})}{\delta 
g^{\mu\nu}}$ the usual energy-momentum tensor, and where the form of 
$H_{\mu\nu}$ is presented  in    Appendix \ref{AppendixA}.

Let us now focus on a cosmological background, namely we consider a flat 
Friedmann-Robertson-Walker (FRW)  background spacetime 
metric of the form
\begin{eqnarray}
ds^2=-dt^2+a(t)^2\d_{ij}dx^i dx^j,
\label{metric}
\end{eqnarray}
where $a(t)$ is the scale factor. 
In this case, the cubic invariant takes the simple form 
  \citep{Erices:2019mkd}  
\begin{equation}\label{24}
P=6\tilde{\beta}H^4\left(2H^2+3\dot{H}\right),
\end{equation}
where  $H=\frac{\dot a}{a}$ is the Hubble parameter and with dots
denoting time derivatives,   
since under the assumptions of being  neither topological
nor trivial and to lead to second-order field equations, it
has only one   free parameter, namely  
\begin{eqnarray}
\tilde{\beta}=-\beta _{1}+4\beta _{2}+2\beta _{3}+8\beta _{4},
\label{basiccond}
\end{eqnarray}
 (note that  the four $\beta_i$ in (6) 
can still be chosen arbitrarily). 
 Moreover, for the matter sector we consider 
  the 
standard perfect fluid,  whose energy-momentum tensor is  
$
T_{\m\n}=(\rho_m+p_m)u_\m u_\n-p_m g_{\m\n}$.
 
The two Friedmann equations of cubic gravity in the case of FRW  metric become
\begin{eqnarray}
\label{Fr11}
3H^2&=&\k\left( \r_m+ \rho_{cub}\right),
\\
3H^2+2\dot{H}&=&-\k\left( p_m +p_{cub}\right),
\label{Fr22}
\end{eqnarray}
where we have defined
\begin{eqnarray}
\label{rholinear}
&&
\r_{cub} \equiv    6\b H^6+\frac{\Lambda}{\k},\\
&&
p_{cub}\equiv  -6  \b  H^4( H^2+2  \dot{H})-\frac{\Lambda}{\k},
\label{plinear}
\end{eqnarray}
and  
where we have merged $\alpha$ and $\tilde{\b}$ in the sole parameter  
$\b\equiv\alpha\tilde{\b}$.  Hence, the cubic terms give rise to an effective 
sector of geometric origin with the above energy density and pressure, and with 
effective  equation-of-state parameter 
\begin{eqnarray}
w_{cub}\equiv  \frac{p_{cub}}{\rho_{cub}}.
\label{wDEdef}
\end{eqnarray}

\section{Perturbation analysis}
\label{Perturbationanalysis}

In the previous section we presented cubic gravity and we extracted the general 
field equations. Additionally, we applied them in a cosmological framework, and 
we provided the background Friedmann equations. Although the Friedmann 
equations do not contain higher-order time-derivatives and thus the theory is 
well-defined at the background level, this does not guarantee that 
instabilities and contradictions with observations will not appear at the 
perturbation level. Hence, in this section we proceed to a 
detailed investigation of the perturbations around a cosmological background. 
As usual, we will investigate the scalar and tensor perturbations separately, 
however we will do that simultaneously since this will give rise to the 
necessary constraints on the model parameters.

\subsection{Scalar perturbations}
 
 Let us start by the examination of scalar perturbations. 
We consider the usual perturbed 
metric of isentropic   perturbations   in the Newtonian gauge  
\cite{Tsujikawa:2007gd,Hwang:2001qk,DeFelice:2011hq,Flender:2012nq,
Basilakos:2013nfa,
Matsumoto:2013sba,Fidler:2017pnb}
\be
ds^2=-a\left(\eta\right)^2
\left(1+2\phi\right) 
d\eta^2+a\left(\eta\right)^2\tensor{\delta}{_i_j}\left(1-2\psi\right)dx^{i}dx^{j
},
\ee
where for convenience we use the conformal time  $\eta$ (with $dt=ad\eta$), 
with $a(\eta)$  the 
corresponding  scale factor, and with $\phi$,$\psi$ the first-order
scalar perturbations.   Furthermore, concerning the perturbations of the matter 
sector, we write  
 \begin{equation}
 \delta T^0_{0}=-\delta\rho,
 \end{equation}
 \begin{equation}
 \delta T^i_{j}=\delta p\delta^i_{j}.
 \end{equation}
 
 Inserting the above into   the general field  equations  (\ref{eom}), and 
transforming as usual to   Fourier space, we find that the time-time 
component of (\ref{eom}) is
 \begin{eqnarray}\label{time}
&&\!\!\!\!\!\!\!\!\!\!\!\!\!\!\!\!\!
2\psi+\frac{\kappa\alpha 
k^2}{a^4}
\Big\{\Big[\left(8\beta_{3}+48\beta_{4}+8\beta_{5}+12\beta_{6}\right.\nonumber\\
&&
\ \ \ \ \  \ \  \ \ \ \,
\left. +40\beta_{7}
+144\beta_{8}\right)\mathcal{H}^2 
\nonumber\\
 &&  \ \ \ \ \ \  \ \ \,
 +\left(48\beta_{2}+32\beta_{3}+80\beta_{4}+20\beta_
{5}\right.
\nonumber\\
 &&\left.\ \ \ \ \ \ \  \ \  \ \ \,
 +24\beta_{6}+56\beta_{7}
+144\beta_{8}\right)\mathcal{H}^\prime 
\Big]\phi \nonumber\\
&& \ \   \ \  \ \
 -\Big[\left(8\beta_{3}+32\beta_{4}+12\beta_{5}+
24\beta_{6}\right.
\nonumber\\
 &&\left. \ \ \ \ \  \ \, \ \  \ \ \,
 +72\beta_{7}
+288\beta_{8}\right)\mathcal{H}^2 \nonumber\\
&& \ \ \ \ \ \,  \ \ \ \ \,
 +\left(12\beta_{1}
+8\beta_{3}+32\beta_{4}+24\beta_{5}+
12\beta_{6}
\right.
\nonumber\\
 &&\left.  \ \ \ \ \  \ \, \ \  \ \ \ \
 +72\beta_{7}
+288\beta_{8}\right)\mathcal{H}^\prime\Big] \psi\Big\}=-\frac{\kappa 
a^2\delta\rho}{k^2},
\end{eqnarray}
with $k$ the wavenumber and where $\mathcal{H}=a^\prime/a$ is the conformal 
Hubble function, with primes denoting conformal-time derivatives. We mention 
that 
since we are interested in  calculating the 
corrections to the gravitational potential, we have kept only the leading terms 
in the $k\gg\mathcal{H}$ regime.

Additionally, the  non-diagonal space-space component equation is found to be
 \begin{eqnarray}
  \label{slip}
  &&\!\!\!\!\!\!\!\!\!\!\!\!\!\!\!\!\!\!\!\!\!\!\!\!\!\!\!\!
  \phi-\psi-\frac{\kappa\alpha
k^2}{a^4}\Big\{\Big[\left(4\beta_{3}+16\beta_{4}+6\beta_{5}\right.\nonumber\\
&&\left. \ \ \ \ \  \   \  
+12\beta_{6}
+36\beta_{7}
+144\beta_{8}\right)\mathcal{H}^2 
 \nonumber\\
&& \ \ \ \ \  \   
 +\left(6\beta_{1}+4\beta_{3}+16\beta_{4}  +12\beta_{5
}\right.\nonumber\\
&&\left.
\ \ \ \ \  \  \
+6\beta_{6}+36\beta_{7}
+144\beta_{8}\right)\mathcal{H}^\prime 
\Big]\phi \nonumber\\ \ \ \ \ \  \ \,  
&& \ \ \  \  
 -\Big[\left(6\beta_{1}+48\beta_{2}+40\beta_{3}+112\beta_{4}\right.\nonumber\\
&& \ \ \ \ \  \ \ \  \left. 
+34\beta_{5}+
36\beta_{6}+100\beta_{7}
+288\beta_{8}\right)\mathcal{H}^2  \nonumber\\
&& \ \ \  \ \ \ \
 +\left(8\beta_{3}
+48\beta_{4}+12\beta_{5}+
18\beta_{6}
\right.\nonumber\\
&& \ \ \ \ \  \ \ \ \ \ \left. 
+68\beta_{7}
+288\beta_{8}\right)\mathcal{H}^\prime\Big] \psi\Big\}=0.
 \end{eqnarray}

\subsection{Tensor perturbations}
 
We continue with the consideration of   tensor perturbations around a flat FRW  
metric, namely we consider
\begin{equation}
 \label{tensor}
ds^2=-a\left(\eta\right)^2 
d\eta^2+a\left(\eta\right)^2\left(\tensor{\delta}{_i_j}+h_{ij}\right)dx^{i}dx^{j
}.
\end{equation}
As usual,  the tensor $h_{ij}$ is 
divergenceless ($\partial^{i}\tensor{h}{_i_j}=0$) 
and traceless ($\tensor{h}{^i_i}=0$). 
The general equation for the tensor perturbations  around a flat 
FRW background, in the case of cubic gravity, whose general field equations are 
given in (\ref{eom}), is given in Eq. (\ref{tensorequation}) in Appendix 
\ref{AppendixB}.

\subsection{Viability conditions}

In the above subsections we examined the scalar and tensor perturbations in 
cubic gravity. Thus, we can now use them in order to extract conditions on the 
model parameters in order for the theory to be healthy and viable.
 In particular, we know that every viable modified gravity is a 
correction on top of general relativity, since the latter must always be  
recovered at a particular limit of the parameters of the modified gravity (in 
our case general relativity is recovered for $\alpha=0$ or equivalently 
$\beta=0$). Obviously, $\beta=0$ is a sufficient condition that no problematic 
features are present, however the goal of the present work is to obtain 
non-trivial versions of the theory, i.e. with 
non-zero $\beta_i$ parameters, that still satisfy the desired requirements, 
namely we want to find   minimal non-zero deviations from general relativity 
that are viable.

A first requirement comes from the  correction in the
Poisson's law. In the case of general 
relativity we have $
\Phi_{eff}\equiv\frac{\phi+\psi}{2} =-\frac{\kappa a^2\delta\rho}{2k^2},
$
which is the quantity that determines the  light bending  
\cite{Keeton:2005jd,Ren:2021uqb}, with 
$\phi=\psi=
 -\frac{\kappa a^2\delta\rho}{2k^2}$. Nevertheless, in general in a modified 
gravity theory the post-Newtonian 
parameter $\gamma\equiv \psi/\phi$ is different than 1, however this deviation 
should be quite small in order to pass the observational tests
 \cite{Will:2014kxa}.

A second requirement is that the tensor perturbations, namely the gravitational 
waves, should propagate at the speed of light $c$, in order to be in agreement 
with LIGO-VIRGO \cite{LIGOScientific:2017vwq} and 
  Fermi Gamma-ray Burst Monitor \cite{Goldstein:2017mmi} observations, which 
require  $|c_{_{GW}}/c-1|\leq 4.5\times 
10^{-16}$ \cite{Monitor:2017mdv}.

Observing the forms of (\ref{time}) and (\ref{slip}) one sufficient condition to 
achieve  $\gamma$ close to one is to choose the model parameters in order to 
make all new terms  apart from one equal to zero (making all of them zero gives 
back general 
relativity).  Furthermore, concerning tensor perturbations,
starting from (\ref{tensorequation})   we  make the standard approximation  
$k^2\sim \mathcal{H}^2$, while we 
consider $\mathcal{H}^\prime\sim\mathcal{H}^2$ and $\tensor{h}{^{(4)}_i_j}\sim 
\mathcal{H}^3\tensor{h}{^{(1)}_i_j}$,  up to $\tensor{h}{^{(1)}_i_j}$ 
term, and we impose the same approximations for the 
$\tensor{h}{_i_j}$ term. Under the above considerations we result to 
 \begin{eqnarray}
&&\beta _{1}=\frac{14}{39}\beta_3+8\beta_4-\frac{34}{39}\beta_5
\nonumber
\\
&&\beta _{2}=-\frac{11}{78}\beta_3-\frac{1}{26}\beta_5\nonumber\\
&&\beta _{6}=\frac{2}{39}\beta_3+8\beta_4-\frac{1}{13}\beta _5\nonumber\\
&&\beta _{7}=-\beta_3-8\beta_4-\frac{1}{2}\beta_5,\nonumber\\
 %\label{conds2a}
  &&\beta 
_{8}=\frac{199}{936}\beta_3+\frac{11}{9}\beta_4+\frac{121}{1404}
\beta_5.
 \label{conds2a2}
\end{eqnarray}
Note that     (\ref{basiccond}) under (\ref{conds2a2}) gives  
$\tilde{\beta}=\frac{14}{13}\beta_3+\frac{28}{39}\beta_5$.
Hence, a theory with  (\ref{conds2a2}), plus the background constraint  
(\ref{basiccond}), corresponds to 
a viable non-trivial minimal deviation from general relativity.

One can clearly see that under conditions   (\ref{conds2a2}), equations  
(\ref{time}) 
and 
(\ref{slip}) provide  the potentials $\phi$ and $\psi$ as 
 \begin{eqnarray}\label{34}
&&\phi=\frac{\left[-2+\frac{7a^4}{14a^4+3k^2\kappa\beta
\left(5\mathcal{H}^2-7\mathcal{H}^{\prime}\right)}\right]\kappa 
a^2\delta\rho}{3k^2}\nonumber\\
&&\psi=-\frac{\left[1+\frac{7a^4}{14a^4+3k^2\kappa\beta
\left(5\mathcal{H}^2-7\mathcal{H}^{\prime}\right)}\right]\kappa 
a^2\delta\rho}{3k^2}.
\end{eqnarray}
 Thus, we do verify that the corrections to the gravitational potentials due 
to cubic terms depend on the single parameter $\beta$ and are minimal, 
satisfying the observed bounds \cite{Will:2014kxa}.
 Hence,  the  post-Newtonian parameter 
$\gamma$ mentioned above becomes
\begin{equation}
\gamma 
\equiv\frac{\psi}{\phi}=\frac{1}{2}+\frac{1}{2-\frac{4}{7}
 \mu\left(2H^2+7\dot { H }
\right)},
\end{equation}
where we have introduced the quantity 
$\mu\equiv\frac{8\pi Gk^2\beta  }{a^2}$ for convenience.
 In the limit $\beta\rightarrow0$ we obtain $\gamma\to 1$ as expected.

Additionally, we can immediately  see that under these conditions the 
gravitational-wave propagation equation becomes
 \begin{equation} 
\tensor{h}{^{\prime\prime}_i_j}+2\mathcal{H}\left(1+\beta_{P}\right)\tensor{h}{^
{\prime}_i_j}+k^2\tensor{h}{_i_j}=0,
\label{tensoreq2}
 \end{equation}
 where
 \begin{equation}
\beta_{P}=-  \frac{3\kappa \beta\left(153\mathcal{H}
^4-488\mathcal{H}^2\mathcal{H}^{
\prime}+235\mathcal{H}^{\prime 2}\right)}{14a^4}.
\label{rel1oeff}
 \end{equation}
 As we observe, under conditions (\ref{conds2a2}), the 
gravitational waves in 
cubic gravity propagate at the speed of light,  and thus the theory is 
viable. However, the cubic terms affect 
the dispersion relation through the term $\beta_{P}$, a feature that appears in 
other viable modified theories of gravity too 
\cite{Ezquiaga:2017ekz,Cai:2018rzd,Bahamonde:2021dqn,Zhu:2021whu}. Lastly, we 
mention 
that 
(\ref{tensoreq2}) can be re-written in terms of   cosmic time, using 
$\mathcal{H}=aH$, 
$\mathcal{H}^{\prime}=a^2(H^2+\dot{H})$    (note that this will bring 
coefficient 
changes between (\ref{rel1oeff}) and  (\ref{rel1oeff2}) below) as
 \begin{equation} 
\tensor{\ddot{h}}{_i_j}+3H\left(1+\beta_P\right)\tensor{\dot{h}}{_i_j}+\frac{k^2
}{a^2}\tensor{h}{_i_j}=0,
 \end{equation}
 where now
 \begin{equation} 
\beta_{P}=\frac{1}{7}\left(100H^4+18H^2\dot{H}-235\dot { H } ^2\right) \kappa 
{\beta}.
\label{rel1oeff2}
 \end{equation}

We proceed by focusing on the scalar perturbations. As we showed, in cubic 
gravity the Poisson's law is modified according to (\ref{34}), which 
implies that we obtain an effective gravitational constant. Using that 
$\kappa=8\pi G$, with $G$ the Newton's constant, we find that the effective 
Newton's constant is given as
\begin{equation}\label{35}
G_{eff}\equiv\frac{1}{3}G\left[4-\frac{7a^4}{7a^4+12\pi 
Gk^2\beta\left(5\mathcal{H}^2-7\mathcal{H}^{\prime}\right)}
\right],
\end{equation}
 in which case one recovers  $\phi=-4\pi G_{eff}  a^2\delta\rho/
k^2$. Note that in terms of cosmic time we have  
\begin{equation}\label{32}
G_{eff}\equiv\frac{1}{3}G\left[4+\frac{1}{\frac{12k^2\pi 
G\beta\left(2H^2+7\dot{H}\right)}{7a^2}-1}\right].
\end{equation}
  Hence, in cubic gravity, as it is typical in modified gravity 
theories, we obtain an effective   Newton's  constant $G_{eff}$ that is in 
general different than $G$,  and the deviation depends on the 
model parameter $\beta$ as well as on     the specific background Hubble 
function evolution, namely on $H(t)$. For  $\beta=0$, in which case cubic 
gravity recovers General Relativity, we obtain $G_{eff}=G$ as expected.   Note 
that although the time-dependence of the effective Newton's 
constant is typical in modified gravity, the scale-dependence appears only in   
subclasses of them \cite{Cai:2015emx,Basilakos:2013nfa}.
According to observations we obtain $G_{eff}/G=1.09 \pm 0.2$ at 1$\sigma$ 
confidence level
\cite{Umezu:2005ee,Nesseris:2006jc,Verbiest:2008gy,Nesseris:2017vor},
and thus this deviation should satisfy  
\begin{equation}
0.89
\lesssim \frac{G_{eff}}{G}  \lesssim1.29.
\label{Geffbounds}
\end{equation}

Now the quantity $\mu\equiv\frac{8\pi Gk^2\beta  }{a^2}$ defined above,  using 
the redshift    $
 1+z=a_0/a$ (with the present scale factor set to $a_0=1$) becomes
\begin{equation}
\mu\left(z\right)=8\pi G {\beta}k^2\left(1+z\right)^2.
\end{equation}
Thus, we can express the effective Newton's constant as
$
G_{eff}=\frac{1}{3}G\left\{4+ [\frac{3}{14} \mu(2H^2+7\dot{H}
)-1]^{-1}\right\}.
$
Therefore, we conclude that if $0< \mu(2H^2+7\dot{H})<14/3$ then we obtain 
$G_{eff}<G$, otherwise $G_{eff}>G$, while for $\mu=0$, i.e. $\beta=0$, we 
recover $G_{eff}=G$.

\begin{figure}[!]
\centering
\hspace{-0.8cm}
\includegraphics[angle=0,width=0.5\textwidth]{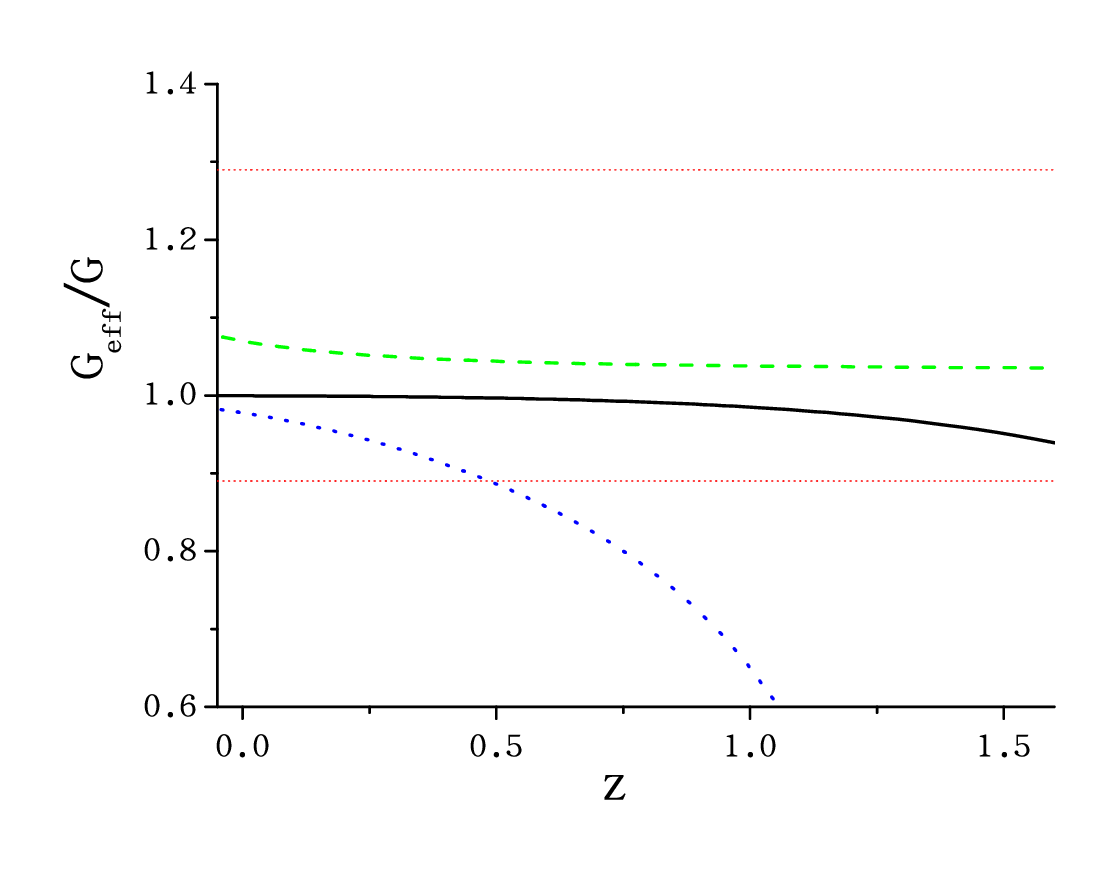}
\caption{ {\it{The evolution of the normalized effective 
gravitational constant  $G_{eff}/G$   as a function of the redshift, at a scale 
$k=10^{-3}$Mpc$^{-1}$, and for $\beta=-0.0001$ (black solid), $\beta=-0.005$ 
(green dashed), and $\beta=0.5$ (blue dotted) in units where $8\pi G=1$ and 
$H_0=1$, where we have set the present matter density parameter 
$\Omega_{m0}\equiv 8\pi G\rho_{m0}/(3H_0^2)\approx 0.31$ 
\cite{Planck:2018vyg}. The horizontal red  dotted lines mark the observational 
bounds on $G_{eff}/G$ 
\cite{Umezu:2005ee,Nesseris:2006jc,Verbiest:2008gy,Nesseris:2017vor} given in 
(\ref{Geffbounds}).
  }}}
  \label{Geffevol}
 \end{figure}
 
\begin{figure}[ht]
\centering
\hspace{-0.8cm}
\includegraphics[angle=0,width=0.5\textwidth]{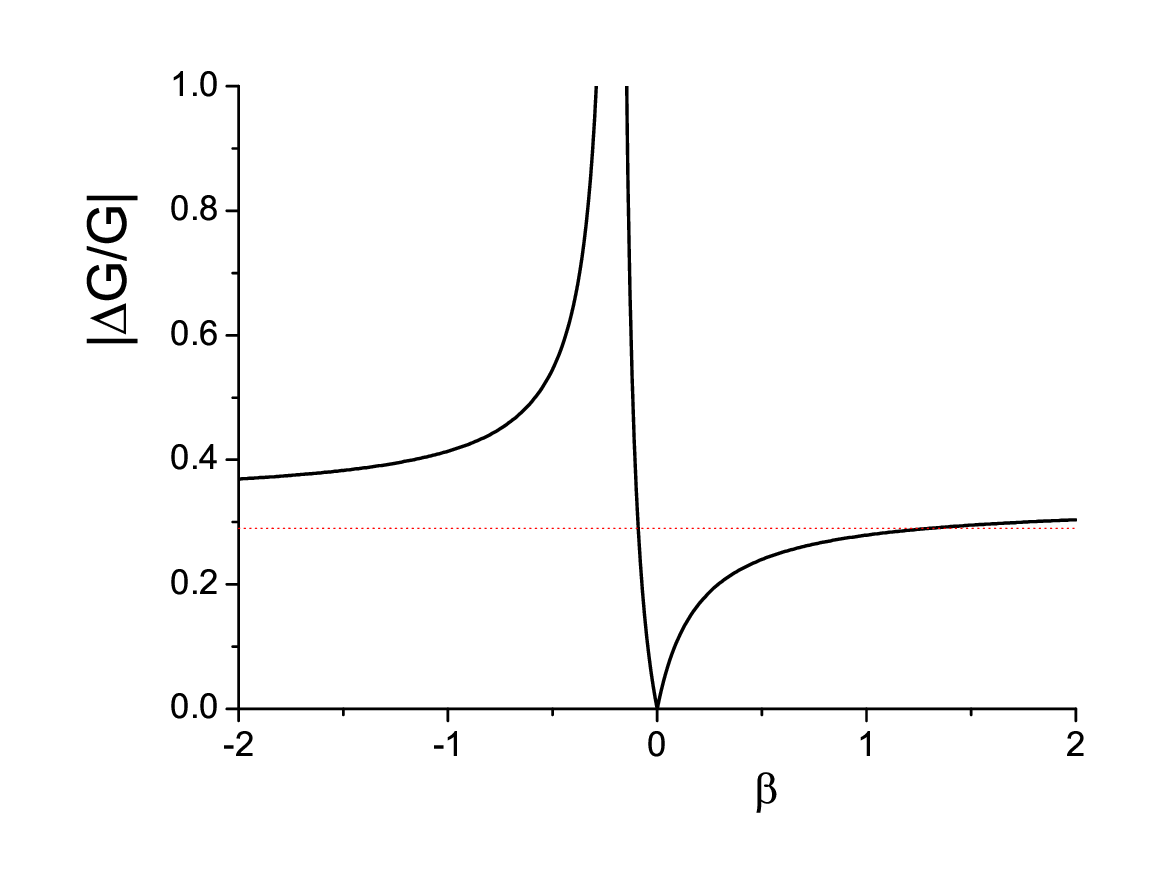}
\caption{ {\it{
The normalized difference of the effective 
gravitational constant  $\left\lvert \Delta G/G\right\rvert$ as a   
function of the model parameter $\beta$, in units where $8\pi G=1$ and 
$H_0=1$, at present time ($a_0=1$ and $\dot{H}(z=0)\approx 
-H_0^2\left(1+q_0\right) $ with $q_{0}\approx-0.503$ the current deceleration 
parameter  \cite{Mamon:2018dxf}). The horizontal red  dotted line marks the 
observational 
bound    $\left\lvert \Delta G/G\right\rvert\lesssim0.29$
\cite{Umezu:2005ee,Nesseris:2006jc,Verbiest:2008gy,Nesseris:2017vor}.  }}
 }
 \label{figA}
 \end{figure}

 In Fig. \ref{Geffevol} we depict the late-time evolution of the 
normalized effective 
gravitational constant  $G_{eff}/G$  as a function of the redshift, in the 
scenario at hand for various choices of $\beta$ in units  $8\pi G=1$ and 
$H_0=1$ (the subscript ``0'' marks the value of a quantity at present), at a 
reference scale $k=10^{-3}$Mpc$^{-1}$, on top of the observational bounds. For  
completeness,
in Fig.  \ref{figA} we depict $\left\lvert\frac{\Delta G}{G}\right\rvert$, 
where $\Delta G\equiv G_{eff}-G$, as a function of the model parameter $\beta$, 
at $k=10^{-3}$Mpc$^{-1}$ at present time  (where $a_0=1$ and 
$\dot{H}(z=0)\approx -H_0^2\left(1+q_0\right) $ with $q_{0}\approx-0.503$ the 
current deceleration parameter  \cite{Mamon:2018dxf}). From these figures we 
deduce that a viable theory should have $-0.01\lesssim\beta\lesssim0.1$. 
Restoring natural units (where $G\approx 6.7\times 10^{-39}$GeV, 
$k=10^{-3}Mpc^{-1}\approx 6.4\times10^{-42}$ GeV, and $H_0=1.4\times 
10^{-42}$GeV) we obtain that $-10^{200}$GeV$^{-2}\lesssim\beta\lesssim 
10^{201}$GeV$^{-2}$. Note that this window is in agreement with the bounds 
obtained  in higher-order corrections to Einstein gravity through causality and 
unitarity considerations on the graviton scattering
\cite{Caron-Huot:2022ugt}.
Lastly, note also that there is a specific 
value of 
$\beta$ in which $G_{eff}$ diverges, as expected from the form of (\ref{32}). 
Finally, we have checked that changing the reference scale $k$ leads to the 
same qualitative features.

  Note that   a varying $G_{eff}$ (of course inside the 
observational bounds), and in particular a $G_{eff}$ smaller than $G$ by a 
suitable amount, is known 
to be one of the mechanisms that can alleviate  the  $H_0$ and $\sigma_8$  
cosmological tensions,   since ``weaker'' gravity can lead to 
faster expansion and smaller matter clustering (see \cite{Abdalla:2022yfr} for 
various models with this property). Thus, the aforementioned property in the 
scenario at hand could be useful towards the tensions alleviation too.

 Let us mention here that the above analysis focuses on late times, while 
at very early times the constraints are typically stronger. If we want to 
extend the analysis up to very early times, namely up to the Big Bang 
Nucleosynthesis (BBN) epoch ($z\sim10^9$), then we should also consider the 
radiation sector, which was neglected above since we focused on late times. 
However, we note that the BBN constraints on cubic gravity were examined in 
\cite{Asimakis:2021yct}. Definitely, in the end of the day, all constraints 
from various investigations should be used simultaneously.

 In summary,   we have extracted the conditions  required for a 
healthy and viable cubic gravity, in order for scalar and tensor perturbations 
to be in agreement with observations, and we extracted the constraints on the 
single parameter $\beta$.

\section{Conclusions}
\label{Conclusions}
 
Cubic gravity is a higher-order modified gravity whose Lagrangian $P$  is 
built by cubic curvature terms under the theoretical requirement to lead to 
second-order field equations at four dimensions. 
Since cubic    and $f(P)$ gravity are known to have interesting cosmological 
phenomenology,  in the present work we investigated the conditions under which 
the theory is healthy and viable at the perturbation level. 

We performed a detailed analysis of the scalar and tensor perturbations. We   
imposed the requirement that 
the two scalar potentials, whose ratio is the   post-Newtonian parameter 
$\gamma$, should   deviate only minimally form the general relativity result. 
Additionally, concerning the tensor perturbations we  imposed the condition 
that the obtained gravitational-wave speed should satisfy the  LIGO-VIRGO 
and Fermi Gamma-ray Burst observations. Thus, we resulted to a 
gravitational-wave equation with gravitational-wave speed equal to the speed of 
light, and where the only deviation from general relativity appears in the 
dispersion relation.

Furthermore, we showed that cubic gravity exhibits an effective Newton's 
constant that depends on the model parameter, on the background evolution, and 
on the wavenumber scale. Hence, by requiring its deviation from the standard 
Newton's constant to be within observational bounds we extracted the 
constraints on the single coupling parameter $\beta$.

In summary, in this work we constructed non-trivial versions of cubic gravity,
namely with non-zero parameters, that satisfy  the viable observational 
requirements. This is a necessary addition to its known  interesting 
cosmological phenomenology.  Clearly, these are not the only classes of 
theories that have this property, since there could be more complicated 
theories, namely  with less constraints and thus more parameters, that share 
this property. Thus, we extracted theories that deviate form general relativity 
at the level of the action, but have minimal deviations (but still non-zero) at 
the level of scalar  and tensor perturbations.

It would be interesting to apply the results of the 
present work in order to study the primordial gravitational waves and the 
primordial black holes in the case of cubic gravity.
 Such an analysis could be useful in order to extract  unique observational 
signatures of cubic gravity, and distinguish this theory from other modified 
theories of gravity.  Additionally, we could
extend the viability investigation in the case of non-linear $f(P)$ gravity and 
examine whether the extra degrees of freedom alter the results, especially those 
related to the gravitational wave propagation.  Furthermore, although 
we have shown that there are no instabilities in our approximated expressions, 
other pathologies could be present, and thus a full  stability analysis of 
tensor perturbations should be performed too. Finally, it would be interesting 
to 
compare the obtained cosmological constraints with constraints arising from 
spherically symmetric solutions.  These studies lie beyond the 
scope of the present work and are left for future projects.

\begin{acknowledgments}

This research is co-financed by Greece and the European Union (European Social 
Fund-ESF) through the Operational Programme ``Human Resources Development, 
Education and Lifelong Learning'' in the context of the project
``Strengthening Human Resources Research Potential via Doctorate Research''
(MIS-5000432), implemented by the State Scholarships Foundation (IKY). 
Additionally, it is supported by Vicerrector\'ia de Investigaci\'on y 
Desarrollo Tecnol\'ogico (Vridt) at Universidad Cat\'olica del Norte through 
N\'ucleo de Investigaci\'on Geometr\'ia Diferencial y Aplicaciones, 
Resoluci\'on Vridt No 096/2022. The authors would like to acknowledge the 
contribution of the COST Action 
CA21136 ``Addressing observational tensions in cosmology with systematics and 
fundamental physics (CosmoVerse)''.

\end{acknowledgments}

\appendix

\section { The general field equations of cubic gravity} 
\label{AppendixA}

Varying the action of cubic gravity (\ref{GB1}) with respect to the metric 
  we obtain the 
general field equations as \cite{Erices:2019mkd}. 

 \begin{equation}\label{eoms2}
G_{\mu\nu}+\Lambda g_{\mu\nu}=\kappa\left(T_{\mu\nu}+\alpha H_{\mu\nu}\right),
\end{equation}
with $T_{\mu\nu}=-\frac{2}{\sqrt{-g}}\frac{\delta(\sqrt{-g}L_{m})}{\delta 
g^{\mu\nu}}$ the   energy-momentum tensor, and where  the 
$\tensor{H}{_\mu_\nu}$ tensor reads as:
\begin{widetext}
\begin{multline}
H_{\mu\nu}=2\left(\beta_1+4\beta_3+40\beta_4-\beta_5+\beta_6\right)g_{\mu\nu}
\tensor{R}{_\alpha^\gamma}R^{\alpha\beta}R_{\beta\gamma}+2\beta_7R_{\alpha\beta}
R^{\alpha\beta}R_{\mu\nu}-2\left(2\beta_3+\beta_5\right)R_{\alpha\beta}\tensor{R
}{_\mu^\alpha}\tensor{R}{_\nu^\beta}\\
\!\!\!\!\!\!\!\!\!\!\!\!\!\!\!\!\!\!\!\!\!\!\!\!\!\!\!\!\!\!\!\!\!\!\!\!\!\!\!\!
\!\!\!\!\!\!\!\!\!\!\!\!\!\!\!\!\!\!\!\!\!\!\!\!\!\!\!\!\!\!\!\!\!\!\!\!\!\!\!\!
\!\!\!\!\!\!\!\!\!\!\!\!\!\!\!\!\!\!\!\!
-\left(\frac{9}{2}
\beta_1+9\beta_3+88\beta_4+\beta_7\right)g_{\mu\nu}R_{\alpha\beta}R^{\alpha\beta
}R_{\mu\nu}-8\beta_4\tensor{R}{_\mu^\alpha}R_{\nu\alpha}R\\
\!\!\!\!\!\!\!\!\!\!\!\!\!\!\!\!\!\!\!\!\!\!\!\!\!\!\!\!\!\!\!\!\!\!\!\!\!\!\!\!
\!\!\!\!\!\!\!\!\!\!\!\!\!\!\!\!\!\!\!\!\!\!\!\!\!\!\!\!\!\!\!\!\!\!\!\!\!\!\!\!
\! 
+6\beta_8R_{\mu\nu}
R^2+2\beta_4R_{\mu\nu}R_{\alpha\beta\gamma\delta}R^{\alpha\beta\gamma\delta}
+\left(\frac{5}{8}\beta_1+\frac{5}{4}\beta_312\beta_4-\beta_8\right)g_{\mu\nu}
R^3\\
\!\!\!\!\!\!\!\!\!\!\!\!\! \! \! \! \! \! \! \! \! 
+\left(3\beta_1+6\beta_3+64\beta_4+\beta
_5-3\beta_6\right)g_{\mu\nu}R^{
\alpha\gamma}R^{\beta\delta}R_{\alpha\beta\gamma\delta}+\left(\frac{8}{3}
\beta_1+\frac{3}{4}\beta_3+7\beta_4\right)g_{\mu\nu}RR_{\alpha\beta\gamma\delta}
R^{\alpha\beta\gamma\delta}\\
-\frac{1}{2}\left[
\beta_1+2\beta_2+3\left(\beta_3+8\beta_4\right)\right]g_{\mu\nu}\tensor{R}{
_\alpha_\beta^\epsilon^\zeta}R^{\alpha\beta\gamma\delta}R_{
\gamma\delta\epsilon\zeta}+4\left(2\beta_4+\beta_7\right)RR^{\alpha\beta}R_{
\mu\alpha\nu\beta}+2\left(2\beta_3+3\beta_6\right)R^{\beta\gamma}\tensor{R}{
_(_\nu^\alpha}\tensor{R}{_\mu_)_\beta_\alpha_\gamma}\\
+\left[
4\left(\beta_3+\beta_6\right)-6\beta_1\right]\tensor{R}{_\alpha^\gamma}R^{
\alpha\beta}R_{\mu\beta\nu\gamma}-\beta_3\tensor{R}{_\nu^\alpha}R_{
\alpha\beta\gamma\delta}\tensor{R}{_\mu^\beta^\gamma^\delta}-24\beta_2\tensor{R}
{_\nu^\alpha}R_{\alpha\gamma\beta\delta}\tensor{R}{_\mu^\beta^\gamma^\delta}
+\left(6\beta_1-4\beta_3+4\beta_5\right)R^{\alpha\beta}R_{
\alpha\gamma\beta\delta}\tensor{R}{_\mu^\gamma_\nu^\delta}\\
\!\!\!\!\!\!\!\!\!\!\!\!\! \! \! \! \! \! \! \! \! \!\!\!\!\!\!\!\!\!\!\!\!\! \! 
\! \! \! \! \! \! \! \!\!\!\!\!\!\!\!\!\!\!\!\! \! \! \! \! \! \! \! \!  \!  \!  
\! 
+4\beta_4R\tensor{R
}{_\mu^\alpha^\beta^\gamma}R_{\nu\alpha\beta\gamma}+6\beta_2R_{
\beta\gamma\delta\epsilon}\tensor{R}{_\mu^\alpha^\beta^\gamma}\tensor{R}{
_\nu_\alpha^\delta^\epsilon}+2\left(6\beta_2+\beta_3\right)R^{\alpha\beta}
\tensor{R}{_\mu_\alpha^\gamma^\delta}R_{\nu\beta\gamma\delta}
\\
 \! \! \! \!\!\!\!\!\!\!\!\!\!\!\!\! \! \! \! \! \! \! 
+2\left(12\beta_2+\beta_3+\beta_5\right)R^{\alpha\beta}\tensor{R}{
_\mu^\gamma_\alpha^\delta}R_{\nu\beta\gamma\delta}
+\left(\beta_3-6\beta_2\right)R_{\alpha\gamma\delta\epsilon}\tensor{R}{
_\mu^\alpha^\beta^\gamma}\tensor{R}{_\nu_\beta^\delta^\epsilon}+9\beta_1R_{
\alpha\delta\gamma\epsilon}\tensor{R}{_\mu^\alpha^\beta^\gamma}\tensor{R}{
_\nu_\beta^\delta^\epsilon}\\
-\beta_3\tensor{R}{_\mu^\alpha}R_{
\alpha\beta\gamma\delta}\tensor{R}{_\nu^\beta^\gamma^\delta}-24\beta_2\tensor{R}
{_\mu^\alpha}R_{\alpha\gamma\beta\delta}\tensor{R}{_\nu^\beta^\gamma^\delta}
+2\left(12\beta_2+\beta_3+\beta_5\right)R^{\alpha\beta}\tensor{R}{
_\mu_\alpha^\gamma^\delta}R_{\nu\gamma\beta\delta}+4\beta_3R^{\alpha\beta}
\tensor{R}{_\mu^\gamma_\alpha^\delta}R_{\nu\gamma\beta\delta}\\
 \! \! \! \!\!\!\!\!\!\!\!\!\!\!\!\! \! \! \! \! \! \! 
 -2\left[
3\beta_1+2\left(\beta_3+\beta_5\right)\right]R^{\alpha\beta}\tensor{R}{
_\mu^\gamma_\alpha^\delta}R_{\nu\delta\beta\gamma}
+\left(\beta_3-6\beta_2\right)R_{\beta\gamma\delta\epsilon}\tensor{R}{
_\mu^\alpha^\beta^\gamma}\tensor{R}{_\nu^\delta_\alpha^\epsilon}+3\beta_1R_{
\beta\delta\gamma\epsilon}\tensor{R}{_\mu^\alpha^\beta^\gamma}\tensor{R}{
_\nu^\delta_\alpha^\epsilon}\\+\left[
6\beta_1+4\left(\beta_3-6\beta_2\right)\right]R_{\alpha\delta\gamma\epsilon}
\tensor{R}{_\mu^\alpha^\beta^\gamma}\tensor{R}{_\nu^\delta_\beta^\epsilon}
+6\beta_1R_{\alpha\epsilon\gamma\delta}\tensor{R}{_\mu^\alpha^\beta^\gamma}
\tensor{R}{_\nu^\delta_\beta^\epsilon}+2\left(4\beta_4+\beta_7\right)R\square 
R_{\mu\nu}+\left(\beta_5+2\beta_7\right)R_{\mu\nu}\square 
R\\
\!\!\!
+\left(\beta_7+12\beta_8\right)g_{\mu\nu}R\square 
R+2\left(\beta_3+8\beta_4+\beta_5+2\beta_7\right)\nabla_{\alpha}R_{\mu\nu}
\nabla^{\alpha}R+\left[\frac{3}{4}\beta_6+2\left(\beta_7+6\beta_8\right)\right]
g_{\mu\nu}\nabla_{\alpha}R_{\mu\nu}\nabla^{\alpha}
R\\
 \! \! \! \!\!\!\!\!\!\!\!\!\!\!
-2\left(\beta_3+\beta_5+\frac{3}{2}\beta_6+2\beta_7\right)R_{\alpha(\nu}
\nabla^{\alpha}\nabla_{\mu)}R+2\left(2\beta_3+\beta_5\right)R^{\alpha\beta}
\nabla_{\beta}\nabla_{\alpha}R_{\mu\nu}+2(2\beta_3+3\beta_6)\tensor{R}{_{(\mu}
^\alpha}\square 
R_{\nu)\alpha}\\+2\left(2\beta_5-2\beta_3-3\beta_6\right)R^{\alpha\beta}\nabla_
{\beta}\nabla_{(\mu}R_{\nu)\alpha}+2\beta_3R^{\alpha\beta\gamma\
delta}\nabla_{\beta}\nabla_{(\mu}R_{\nu)\alpha\gamma\delta}
-2\left(3\beta_1+12\beta_2+2\beta_3+\beta_5\right)\nabla_{\alpha}R_{\nu\beta}
\nabla^{\beta}\tensor{R}{_\mu^\alpha}
\\+\left(24\beta_2+8\beta_3+6\beta_6\right)\nabla_{\beta}R_{\nu\alpha}\nabla^{
\beta}\tensor{R}{_\mu^\alpha}+\left(2\beta_7+3\beta_6-\beta_5\right)g_{\mu\nu}R_
{\alpha\beta}\nabla^{\beta}\nabla^{\alpha}
R+\left(8\beta_4+2\beta_3-3\beta_1\right)R_{\mu\alpha\nu\beta}\nabla^{\beta}
\nabla^{\alpha}R\\+2\left(\beta_5+2\beta_7\right)g_{\mu\nu}R^{\alpha\beta}
\square 
R_{\alpha\beta}+2\left(3\beta_1+\beta_5\right)R_{\mu\alpha\nu\beta}\square 
R^{\alpha\beta}+2\beta_5R^{\alpha\beta}\square 
R_{\mu\alpha\nu\beta}+\left(3\beta_6-4\beta_5-2\beta_3\right)g_{\mu\nu}\nabla_{
\beta}R_{\alpha\gamma}\nabla^{\gamma}R^{\alpha\beta}\\- 
2\left(2\beta_3-6\beta_1\right)\nabla_{\beta}R_{\alpha(\mu\nu)\gamma}\nabla^{
\gamma}R^{\alpha\beta}+2\left[\beta_3+2\left(\beta_5+\beta_7\right)\right]g_{
\mu\nu}\nabla_{\gamma}R_{\alpha\beta}\nabla^{\gamma}R^{\alpha\beta}
+4\left(3\beta_1+\beta_5\right)\nabla_{\gamma}R_{\mu\alpha\nu\beta}\nabla^{
\gamma}R^{\alpha\beta}\\-2\left(\beta_3-6\beta_1\right)\nabla_{\beta}R_{
\alpha(\mu\nu)\gamma}\nabla^{\gamma}R^{\alpha\beta}+2\left[
\beta_3+2\left(\beta_5+\beta_7\right)\right]g_{\mu\nu}\nabla_{\gamma}R_{
\alpha\beta}\nabla^{\gamma}R^{\alpha\beta}+4\left(3\beta_1+\beta_5\right)\nabla_
{\gamma}R_{\mu\alpha\nu\beta}\nabla^{\gamma}R^{\alpha\beta}
\\-2\left(\beta_3-6\beta_1\right)R_{\beta\gamma\alpha(\nu}\nabla^{\gamma}
\nabla^{\beta}\tensor{R}{_{\mu)}^\alpha}-4\left(12\beta_2+\beta_3\right)R_{
\alpha\gamma\beta\nu}\nabla^{\gamma}\nabla^{\beta}\tensor{R}{_\mu^\alpha}
-4\left(3\beta_1+\beta_3+\beta_5\right)R_{\beta\alpha\gamma(\nu}
\nabla^\gamma\nabla_{\mu)}R^{\alpha\beta}\\
\!\!\!\!\!\!\!\!\!\!\!\!\!\!\!\!\!\!\!\!\!\!\!
+6\beta_1R^{\alpha\beta\gamma\delta}
\nabla_{\delta}\nabla_{\beta}R_{\mu\alpha\nu\gamma}+2\beta_3\tensor{R}{_{(\nu}
^\alpha^\beta^\gamma}R_{\mu)\alpha\beta\gamma}+12\beta_2\nabla_{\alpha}R_{
\nu\delta\beta\gamma}\nabla^{\delta}\tensor{R}{_\mu^\alpha^\beta^\gamma}
-6\beta_1\nabla_{\gamma}R_{\nu\beta\alpha\delta}\nabla^{\delta}\tensor{R}{
_\mu^\alpha^\beta^\gamma}\\
\!\!\!
\!\!\! +2\beta_3\nabla_{\delta}R_{\nu\alpha\beta\gamma}
\nabla^{\delta}\tensor{R}{_\mu^\alpha^\beta^\gamma}
+2\left(2\beta_3+8\beta_4+\beta_5\right)g_{\mu\nu}R_{\alpha\gamma\beta\delta}
\nabla^{\delta}\nabla^{\gamma}R^{\alpha\beta}+\frac{1}{2}
\left(\beta_3+8\beta_4\right)g_{\mu\nu}\nabla_{\epsilon}R_{
\alpha\beta\gamma\delta}\nabla^{\epsilon}R^{\alpha\beta\gamma\delta}
\\+\left(6\beta_1-2\beta_3+2\beta_5-3\beta_6\right)\nabla^{\beta}\tensor{R}{
_\nu^\alpha}\nabla_{\mu}R_{\alpha\beta}-2\left(\beta_3+8\beta_4+\frac{3}{2}
\beta_{6}+2\beta_7\right)\nabla^{\alpha}R\nabla_{(\mu}R_{\nu)\alpha}
+\beta_3\nabla^{\delta}\tensor{R}{_\nu^\alpha^\beta^\gamma}\nabla_{\mu}R_{
\alpha\delta\beta\gamma}\\+4\left(\beta_3+\beta_5\right)\nabla^{\gamma}R^{
\alpha\beta}\nabla_{(\mu}R_{\nu)\alpha\beta\gamma}
+\left(6\beta_1-2\beta_3+2\beta_5-3\beta_6\right)\nabla^{\beta}\tensor{R}{
_\mu^\alpha}\nabla_{\nu}R_{\alpha\beta}+2\left[
3\beta_1+2\left(\beta_5+\beta_7\right)\right]\nabla_{\mu}R^{\alpha\beta}\nabla_{
\nu}R_{\alpha\beta}\\-\left[\frac{1}{2}
\beta_5+2\left(\beta_7+6\beta_8\right)\right]\nabla_{\mu}R\nabla_{\nu}
R-4\beta_4\nabla_{\mu}R^{\alpha\beta\gamma\delta}\nabla_{\nu}R_{
\alpha\beta\gamma\delta}+\beta_3\nabla^{\delta}\tensor{R}{
_\mu^\alpha^\beta^\gamma}\nabla_{\nu}R_{\alpha\delta\beta\gamma}
-4\left(\beta_5+\beta_7\right)R^{\alpha\beta}\nabla_{\nu}\nabla_{\mu}R_{
\alpha\beta}\\
\!\!\!\!\!\!\!\!\!\!\!\!\!\!\!\!\!\!\!\!\!\!\!\!\!\!\!\!\!\!\!\!\!\!\!\!\!\!\!\!
\!\!\!\!\!\!\!\!\!\!\!\!\!\!\!\!\!\!\!\!\!\!\!\!\!\!\!\!\!\!\!\!\!\!\!\!\!\!\!\!
\!\!\!\!\!\!\!\!\!\!\!\!\!\!\!\!\!\!\!\!\!\!\!\!\!\!\!\!\!\!\!
-2\left(2\beta_4+\beta_7+6\beta_8\right)R\nabla_{\nu}\nabla_{\mu}
R-4\beta_4R^{\alpha\beta\gamma\delta}\nabla_{\nu}\nabla_{\mu}R_{
\alpha\beta\gamma\delta}. \ \ \ \ \ \ \ \ \ \ \ \ \ \ \ \ \ \ \ \ \ \ \ \ \ \ \ 
\ \ \ \ \ \ \ \ \ \ \ \ \ \ \ \ 
\ \ \ \ \ \ \ \ \ \ \ \ \ \ \ \ \ \ \ \ \ \ \ 
\end{multline}
\end{widetext}

\section { The general equation of tensor perturbations} 
\label{AppendixB}

The general equation for the tensor perturbations (\ref{tensor}) around a flat 
FRW background, in the case of cubic gravity whose general field equations are 
given by (\ref{eoms2}), is given by  
\begin{widetext}
 \begin{multline}\!
-2\alpha\kappa\left\{\left[3\beta_{1}-4\beta_{3}-24\beta_{4}+\beta_{5}
-6\left(\beta_{6}+\beta_{7}\right)\right]\mathcal{H}^2-3\left(8\beta_{2}+4\beta_
{3}+8\beta_{4}+2\beta_{5}+\beta_{6}+2\beta_{7}\right)\mathcal{H}^\prime\right\}
\tensor{h}{^{(4)}_i_j}\\
\!\!\!\!\!\!\!\!\!\!\!\!\!\!\!\!\!\!\!\!\!\!\!\!\!\!\!\!\!\!\!\!\!\!\!\!\!\!\!\!
\!\!\!\!\!\!\!\!\!\!\!\!
-4\alpha\kappa\left\{2\left[-3\beta_{1}+4\beta_{3}
+24\beta_{4}-\beta_{5}+6\left(\beta_{6}+\beta_{7}\right)\right]\mathcal{H}
^3+2\left(3\beta_{1}+24\beta_{2}+8\beta_{3}+7\beta_{5}-3\beta_{6}\right)\mathcal
{H}\mathcal{H}^\prime\right.\\
\left.
\!\!\!\!\!\!\!\!\!\!\!\!\!\!\!\!\!\!\!\!\!\!\!\!\!\!\!\!\!\!\!\!\!\!\!\!\!\!\!\!
\!\!\!\!\!\!\!\!\!\!\!\!\!\!\!\!\!\!\!\!\!\!\!\!\!\!\!\!\!\!\!\!\!\!\!\!\!\!\!\!
\!\!\!\!\!\!\!\!\!\!\!\!\!\!\!\!
\!\!\!\!\!\!\!\!\!\!\!\!\!\!\!
\!\!\!\!\!\!\!\!\!\!\!\!\!\!\! -3\left(8\beta_{2}+4\beta_{3}+8\beta_{4}
+2\beta_{5}+\beta_{6}+2\beta_{7}\right)\mathcal{H}^{\prime\prime}\right\}\tensor
{h}{^{(3)}_i_j}\\
\!\!\!\!\!\!\!\!\!\!\!\!\!\!\!
\!\!\!\!\!\!\!\!\!\!\!\!\!\!\!\!\!\!\!\!\!\!\!\!\!\!\!\!\!\!\!\!\!\!\!\!\!\!\!\!
\!\!\!\!\!\!\!\!\!\!\!\!\!\!\!\!\!\!\!\!\!\!\!\!\!\!\!\!\!\!\!\!\!\!\!
\!\!\!\!\!\!\!\!\!\!\!\!\!\!\!
\!\!\!\!\!\!\!\!\! +\left\{ 
a^4+2\alpha\kappa\left\{3\beta_{1} -4\left(3\beta_{2}
+2\beta_{3}-3\beta_{4}+\beta_{5}+6\beta_{6}+9\beta_{7}+27\beta_{8}
\right)\mathcal{H}^4\right.\right.\\
\!\!\!\!\!\!\!\!\!\!\!\!\!\!\!\!\!\!\!\!\!\!\!\!\!\!\!\!\!\!\!\!\!\!\!\!\!\!\!\!
\!\!\!\!\!\!\!\!\!\!\!\!\!\!\!\!\!\!\!\!\!\!\!\!\!\!\!\!\!\!\!\!\!\!\!\!\!\!\!\!
\!\!\!\!\!\!\!\!\!\!\!\!\!\!\!\!\!\!\! 
\left.\left.-\left(3\beta_{1}-24\beta_{2}+26\beta_{3}+192\beta_{4}+22\beta_{5}
+36\beta_{6}+96\beta_{7}+216\beta_{8}\right)\mathcal{H}^2 
\mathcal{H}^\prime\right.\right.\\\left.\left.-3\left(16\beta_{2}+8\beta_{3}
+12\beta_{4}+5\beta_{5}+5\beta_{6}+12\beta_{7}+36\beta_{8}\right)\mathcal{H}^{
\prime 
2}+2k^2\left\{\left[-3\beta_{1}+12\beta_{2}+8\beta_{3}+24\beta_{4}+\beta_{5}
+6\left(\beta_{6}+\beta_{7}\right)\right]\mathcal{H}
^2\right.\right.\right.\\\left.\left.\left.+\left(12\beta_{2}+8\beta_{3}
+24\beta_{4}+4\beta_{5}+3\beta_{6}+6\beta_{7}\right)\mathcal{H}^\prime\right\}
-\left(3\beta_{1}+72\beta_{2}+26\beta_{3}+24\beta_{4}+16\beta_{5}-6\beta_{6}
\right)\mathcal{H}\mathcal{H}^{\prime\prime}
\right.\right.\\\left.\left.+3\left(8\beta_{2}+4\beta_{3}+8\beta_{4}+2\beta_{5}
+\beta_{6}+2\beta_{7}\right)\mathcal{H}^{\prime\prime\prime}\right\}\right\}
\tensor{h}{^\prime^\prime_i_j}+\left\{2a^4\mathcal{H}-2\alpha\kappa\left\{6\left
[5\beta_{1}-4\left(\beta_{2}+2\beta_{3}+7\beta_{4}+4\beta_{6}+5\beta_{7}+9\beta_
{8}\right)\right]\mathcal{H}^5\right.\right.\\
\!\!\!\!\!\!\!\!\!\!\!\!\!\!\!\!\!\!\!\!\!\!\!\!\!\!\!\!\!\!\!\!\!\!\!\!\!\!\!\!
\!\!\!\!\!\!\!\!\!\!\!\!\!\!\!\!\!\!\!\!\!\!\!\!\!\!\!\!\!\!\!\!\!\!\!\!\!\!\!\!
\!\!\!\!\!\!\!\!\!\!\!\!\!\!\!\!\!\!\!\!\!\!\!\!\!\!\!\!\!\!
\left.\left.+2\left[-63\beta_{1}+2\left(-24\beta_{2}+7\beta_{3}+60\beta_{4}-
28\beta_{5}+54\beta_{6}+30\beta_{7}\right)\right]\mathcal{H}^3\mathcal{H}^{
\prime}\right.\right.\\
\!\!\!\!\!\!\!\!\!\!\!\!\!\!\!\!\!\!\!\!\!\!\!\!\!\!\!\!\!\!\!\!\!\!\!\!\!\!\!\!
\!\!\!\!\!\!\!\!\!\!\!\!\!\!\!\!\!\!\!\!\!\!\!\!\!\!\!\!\!\!\!\!\!\!\!\!\!\!\!\!
\!\!\!\!\!\!\!\!\!\!\!\!\!\!\!\!\!\!\!\!\!\!
\left.\left.+2\left(39\beta_{1}+72\beta_{2}+26\beta_{3}+12\beta_{4}+55\beta_{5}-
36\beta_{6}+24\beta_{7}+108\beta_{8}\right)\mathcal{H}\mathcal{H}^{\prime 
2}
\right.\right.\\\!\!\!\!\!\!\!\!\!\!\!\!\!\!\!\!\!\!\!\!\!\!\!\!\!\!\!\!\!\!\!\!
\!\!\!\!\!\!\!\!\!\!\!\!\!\!\!\!\!\!\!\!\!\!\!\!\!\!\!\!\!\!\!\!\!\!\!\!\!\!\!\!
\!\!\!\!\!\!\!\!\!\!\!\!\!\!\!\!\!\!\!\!\!\!\!\!\!\!\!\!\!\!\!\!\!\!\!\!\!\!\!\!
\!\!\!\!\!\!\!
\left.\left.+\left(39\beta_{1}+120\beta_{2}+62\beta_{3}+96\beta_{4}+76\beta_{5}
+84\beta_{7}+216\beta_{8}\right)\mathcal{H}^2\mathcal{H}^{\prime\prime}
\right.\right.\\
\!\!\!\!\!\!\!\!\!\!\!\!\!\!\!\!\!\!\!\!\!\!\!\!\!\!\!\!\!\!\!\!\!\!\!\!\!\!\!\!
\!\!\!\!\!\!\!\!\!\!\!\!\!\!\!\!\!\!\!\!\!\!\!\!\!\!\!\!\!\!\!\!\!\!\!\!\!\!\!\!
\!\!\!\!\!\!\!\!\!\!\!\!\!\!\!\!\!\!\!\!\!\!\! \! \! 
\left.\left.+\left(-15\beta_{1}-24\beta_{2}+2\beta_{3}+48\beta_{4}-8\beta_{5}
+36\beta{6}+60\beta_{7}+216\beta_{8}\right)\mathcal{H}^\prime\mathcal{H}^{
\prime\prime}\right.\right.\\\!\!\!\!\!\!\!\!\!\!\!\!\!\!\!\!\!\!\!\!\!\!\! \! 
\! \!\!\!\!\!\!\!\!\!\!\!\!\!\!\!\!\!\!\!\!\!\!\! \! \! 
\!\!\!\!\!\!\!\!\!\!\!\!\!\!\!\!\!\!\!\! \!
\left.\left.-k^2\left\{4\left[3\beta_{1}-12\beta_{2}-8\beta_{3}-24\beta_{4}
-\beta_{5}
-6\left(\beta{6}+\beta_{7}\right)\right]\mathcal{H}^3-12\left(\beta_{1}+\beta_{5
}-\beta_{6}\right)\mathcal{H}\mathcal{H}^{\prime}
\right.\right.\right.\\
\! \!\!\!\!\!\!\!\!\!\!\!\!\!\!\!\!\!\!\!\!\!\!\! \! \! 
\!\!\!\!\!\!\!\!\!\!\!\!\!\!\!\!\!\!\!\! \!\! 
\!\!\!\!\!\!\!\!\!\!\!\!\!\!\!\!\!\!\!\!\!\!\! \! \! 
\!\!\!\!\!\!\!\!\!\!\!\!\!\!\!\!\!\!\!\! \!\! 
\!\!\!\!\!\!\!\!\!\!\!\!\!\!\!\!\!\!\!\!\!\!\! \! \! 
\!\!\!\!\!\!\!\!\!  
\left.\left.\left.+2\left(12\beta_{2}+8\beta_{3}+24\beta_{4}+4\beta_{5}
+3\beta{6}+06\beta_{7}\right)\mathcal{H}^{\prime\prime}\right.\right.\right.\\
\!\!\!\!\!\!\!\!\!\!\!\!\!\!\!\!\!\!\!\!\!\!\! \! \! 
\!\!\!\!\!\!\!\!\!\!\!\!\!\!\!\!\!\!\!\! \!\! 
\!\!\!\!\!\!\!\!\!\!\!\!\!\!\!\!\!\!\!\!\!\!\! \! \! 
\!\!\!\!\!\!\!\!\! \!\!\!\!\!\!\! 
\left.\left.\left.-\left[3\beta_{1}+2\left(12\beta_{2}+7\beta_{3}+12\beta_{4}
+5\beta_{5}
+3\beta{6}+6\beta_{7}\right)\right]\mathcal{H}\mathcal{H}^{\prime\prime\prime}
\right\}\right\}\right\}\tensor{h}{^\prime_i_j}\\
\!\!\!\!\!\!\!\!\!  
\!\!\!\!\!\!\!\!\!   
\!\!\!\!\!\!\!\!\!    
\!\!\!\!\!\!\!\!\!    
\!\!\!\!\!  +k^2\left\{-2\alpha\kappa 
k^2\left\{\left[3\beta_{1}-10\beta_{3}-3\left[8\beta_{4}+\beta_{5}+2\left(\beta_
{6}+2\beta_{7}\right)\right]\right]\mathcal{H}^2  
-2\left(4\beta_{3}+24\beta_{4}+2\beta_{5}+3\beta_{6}+6\beta_{7}
\right)\mathcal{H}^\prime\right\}\right.
\\  
 \!\!\!\!\!\!\!\!\!   \!\!\!\!\!\!\!\!\!   \!\!\!\!\!\!\!\!\!   
\!\!\!\!\!\!\!\!\!   \!\!\!\!\!\!\!\!\!   \!\!\!\!\!\!\!\!\!   
\!\!\!\!\!\!\!\!\!   \!\!\!\!\!\!\!\!\!   \!\!\!\!\!\!\!\!\!   
\!\!\!\!\!\!\!\!\!   \!\!\!\!\!\!\!\!\!   \!\!\!\!\!\!\!\!\!\!\!   
 +a^4-2\alpha\kappa\left\{3\left[3\beta_{1}
-4\left(3\beta_{2}+2\beta_{3}+\beta_{4}+3\beta_{6}+\beta_{7}-9\beta_{8}
\right)\right]\mathcal{H}^4\right. \\
\!\!\!\!\!\!\!\!\!  \!\!\!\!\!\!\!\!\!  \!\!\!\!\!\!\!\!\!  
\!\!\!\!\!\!\!\!\!  \!\!\!\!\!\!\!\!\!  \!\!\!\!\!\!\!\!\!  
\!\!\!\!\!\!\!\!\!  \!\!\!\!\!\! 
\left.\left.-\left[27\beta_{1}-2\left(36\beta_{2}+23\beta_{3}+24\beta_{4}+\beta_
{5}+42\beta_{6}+48\beta_{7}+108\beta_{8}\right)\right]\mathcal{H}^2\mathcal{H}
^\prime\right.\right.\\
\left.\left.+\left[(12\beta_{1}+8\beta_{3}+36\beta_{4}+19\beta_{5}+3\beta_{6}+
36\left(\beta_{7}+3\beta_{8}\right)\right]\mathcal{H}^{\prime 
2}+\left(15\beta_{1}+10\beta_{3}+24\beta_{4}+20\beta_{5}+6\beta_{6}+
24\beta_{7}\right)\mathcal{H}\mathcal{H}^{\prime\prime}\right.\right.\\  
\left.\left.-\left(3\beta_{1}+2\beta_{3}+4\beta_{5}+3\beta_{6}+
6\beta_{7}\right)\mathcal{H}^{\prime\prime\prime}\right\}\right\}\tensor{h}{_i_j
}=0.\  \,
\label{tensorequation}
\end{multline}
\end{widetext}

\end{document}